\documentclass[twocolumn]{aastex631}
\usepackage{array}
\usepackage{float}
\usepackage{xcolor}
\usepackage{multirow}
\usepackage{amssymb}
\usepackage{CJKutf8}
\usepackage{cancel}

\begin{document}

\title{Formation of Recycled Pulsars in Common Envelope Binaries}

\author[0009-0009-4482-6350]{Yu-Dong Nie}
\affiliation{Department of Astronomy, Nanjing University, Nanjing 210023, People's Republic of China}
\affiliation{Key Laboratory of Modern Astronomy and Astrophysics, Nanjing University, Ministry of Education, Nanjing 210023, People's Republic of China}

\author[0000-0003-2506-6906]{Yong Shao}
\email{shaoyong@nju.edu.cn}
\affiliation{Department of Astronomy, Nanjing University, Nanjing 210023, People's Republic of China}
\affiliation{Key Laboratory of Modern Astronomy and Astrophysics, Nanjing University, Ministry of Education, Nanjing 210023, People's Republic of China}

\author[0000-0003-3862-0726]{Jian-Guo He}
\affiliation{Department of Astronomy, Nanjing University, Nanjing 210023, People's Republic of China}
\affiliation{Key Laboratory of Modern Astronomy and Astrophysics, Nanjing University, Ministry of Education, Nanjing 210023, People's Republic of China}

\author[0009-0001-4454-8428]{Ze-Lin Wei}
\affiliation{Department of Astronomy, Nanjing University, Nanjing 210023, People's Republic of China}
\affiliation{Key Laboratory of Modern Astronomy and Astrophysics, Nanjing University, Ministry of Education, Nanjing 210023, People's Republic of China}

\author[0000-0002-0822-0337]{Shi-Jie Gao}
\affiliation{Department of Astronomy, Nanjing University, Nanjing 210023, People's Republic of China}
\affiliation{Key Laboratory of Modern Astronomy and Astrophysics, Nanjing University, Ministry of Education, Nanjing 210023, People's Republic of China}

\author[0000-0002-3614-1070]{Xiao-Jie Xu}
\affiliation{Department of Astronomy, Nanjing University, Nanjing 210023, People's Republic of China}
\affiliation{Key Laboratory of Modern Astronomy and Astrophysics, Nanjing University, Ministry of Education, Nanjing 210023, People's Republic of China}

\author[0000-0002-0584-8145]{Xiang-Dong Li}
\affiliation{Department of Astronomy, Nanjing University, Nanjing 210023, People's Republic of China}
\affiliation{Key Laboratory of Modern Astronomy and Astrophysics, Nanjing University, Ministry of Education, Nanjing 210023, People's Republic of China}




\begin{abstract}


We present a systematic study of the evolution of low- and intermediate-mass X-ray binaries (L/IMXBs) consisting of a $1.4\,M_{\odot}$ neutron star (NS) and a donor star of mass $1-8\,M_{\odot}$. Using grids of detailed MESA simulations, we show that for donor masses of $2-8\,M_{\odot}$, mass transfer may be dynamically unstable, leading to a common envelope (CE) phase. By adopting CE ejection efficiencies in the range $\alpha_{\rm CE} = 0.3-3.0$, we find that post-CE binaries frequently experience a CE decoupling phase (CEDP), which plays a critical role in determining their final orbital and compositional properties. Systems with initial donor masses $\gtrsim 3.5\,M_{\odot}$ predominantly evolve into NS binaries with carbon-oxygen or oxygen-neon white dwarfs (WDs) with masses between $0.5\,M_{\odot}$ and $1.4\,M_{\odot}$. Comparison with the observed population of binary pulsars with a WD companion shows better agreement with higher CE ejection efficiencies ($\alpha_{\rm CE} = 3.0$). Furthermore, we demonstrate that NSs can accrete a sufficient amount of matter ($\gtrsim 0.01\,M_{\odot}$) during the CEDP and subsequent Case BA/BB/BC mass transfer phases to be effectively recycled into millisecond pulsars. We identify two distinct evolutionary channels capable of reproducing the observed characteristics of the millisecond pulsar PSR J1928+1815 with a helium-star companion.  Our results highlight the importance of the CEDP in the formation of recycled pulsars and provide constraints on the CE ejection efficiency during binary evolution.

\end{abstract}


\section{Introduction} \label{sec:introduction}

Pulsars, the observational manifestations of neutron stars (NSs), have been a subject of intense study since the first radio pulsar was discovered by \citet{Hewish1968}. 
To date, over 3000 radio pulsars have been detected in the Milky Way, among which more than 200 are recycled pulsars residing in binary systems. This indicates that binary evolution plays a significant role in the formation of recycled pulsars. Systems containing a recycled pulsar and a white dwarf (WD) are widely considered to be the evolutionary endpoints of low- and intermediate-mass X-ray binaries \citep[L/IMXBs,][]{Tauris2023}. Approximately 200 LMXBs have been observed in our Galaxy \citep{Liu2007,Chaty2022abn..book.....C}, primarily located in the Galactic bulge and globular clusters \citep[e.g.,][]{Paradijs1995}. In contrast, only a handful of IMXBs have been identified, largely due to their relatively short-lived mass transfer (MT) phases ($\sim 1000\,$yr) and the absorption of X-rays by the dense gas surrounding the accreting NSs \citep{Heuvel1975}. Investigating the evolution of IMXBs is essential for understanding the origin of binary pulsars with massive carbon-oxygen (CO) or oxygen-neon (ONe) WD companions. In L/IMXBs, the accretion of mass and angular momentum can spin up NSs into rapidly rotating pulsars via the so-called recycling process \citep[e.g.,][]{Tauris2023}.


The canonical formation channels for binaries hosting a recycled pulsar are well established. An initial binary system consisting of a high-mass primary and a low-mass secondary may undergo a common envelope (CE) phase, provided the primary is sufficiently massive to end its life as an NS. Following the CE phase, the naked helium (He) star may engage in further MT phase. If the binary remains bound after the primary undergoes a supernova explosion, an LMXB forms. Close-orbit LMXBs lose angular momentum via magnetic braking and gravitational wave radiation, causing their orbits to shrink. If the donor star is hydrogen rich and always remains Roche-lobe filling, the binary evolves to become a converging system \citep{Pylyser1988,Istrate2014}. Additionally, LMXBs with a narrow range of initial orbital periods can evolve into binaries containing a low-mass HeWD \citep{Sluys2005,Istrate2014}. This outcome is sensitive to input physics, particularly the treatment of magnetic braking \citep[e.g.,][]{Istrate2014,Van2019,Deng2021,Chen2021MNRAS.503.3540C,Gao2022,Fan2024}. Furthermore, the formation of recycled pulsar$-$HeWD binaries has also been linked to wide-orbit LMXBs \citep{Rappaport1995,Tauris1999,Podsiadlowski2002, Lin2011,Shao2012,Jia2016,Gao2023MNRAS.525.2605G,Wang2024MNRAS.532.2196W}.

Analogous to LMXBs, an initial binary with a high-mass primary and an intermediate-mass secondary can evolve into an IMXB if it survives the supernova explosion associated with NS formation. Close-orbit IMXBs typically undergo Case A MT (where Roche-lobe overflow, RLOF, occurs during the donor's core hydrogen burning) or early Case B MT (RLOF during the Hertzsprung gap phase), ultimately yielding a binary with a recycled pulsar and a HeWD or a COWD companion \citep{Tauris2000a,Shao2012}. For wider-orbit IMXBs, RLOF commences when the donor ascends the giant branch or a later evolutionary stage. This MT phase is often dynamically unstable, leading to a CE phase during which the NS becomes engulfed by the donor's envelope \citep{Paczynski1976,VandenHeuvel1976}. A recycled pulsar with a COWD companion can form from such a CE-evolving IMXB \citep{vandenHeuvel1994}, highlighting the pivotal role of CE evolution in IMXB outcomes. 

The CE evolutionary scenario, as proposed by \citet{Paczynski1976} and reviewed by \citet{Ivanova2013}, outlines two critical aspects: the initial conditions for triggering CE evolution related to MT stability \citep[e.g.,][]{Soberman1997,Ge2010,Ge2015,Ge2020,Pavlovskii2017,Han2020,Shao2014ApJ...796...37S,Shao2021,Marchant2021} and the final outcomes related to the balance between orbital decay and envelope ejection \citep[e.g.,][]{Webbink1984,Nelemans2005,Soker2015,Klencki2021,VG2022,Hirai2022,DiStefano2023}. Although binary pulsar formation channels have been studied for decades, the detailed impact of CE evolution remains an active area of research. Previous works \citep[e.g.,][]{Tauris2011b,Zhu2015,He2024} often relied on analytic parameterizations for CE outcomes, underscoring the need for detailed simulations to accurately determine CE properties. By modeling the evolution of high-mass X-ray binaries through a CE phase, \citet{Nie2025} recently identified a new post-CE evolutionary stage—the common envelope decoupling phase (CEDP)—providing a valuable framework for probing CE and post-CE evolution. 

Using the updated CE scheme of \citet{Marchant2021}, \citet{Nie2025} performed simulations for binary systems containing a $1.4\,M_{\odot}$ NS and an $8-20\,M_{\odot}$ donor star, following evolution until core carbon depletion. In this paper, we similarly simulate a grid of NS L/IMXBs, focusing on systems that undergo CE phases, and compare our theoretical results with observational data from binary pulsars. 

\section{Method} \label{sec:method}

We employ the Modules for Experiments in Stellar Astrophysics (MESA) code \citep[version 12115,][]{Paxton2011,Paxton2013,Paxton2015,Paxton2018,Paxton2019,Jermyn2023} to conduct our binary evolution simulations. Each  model is initialized with a zero-age main-sequence (ZAMS) donor star and an NS of mass $M^{\rm i}_{\rm NS}=1.4\,M_{\odot}$.
We construct a comprehensive grid of models spanning initial donor masses $M^{\rm i}_{\rm d}$ from 1\,$M_{\odot}$ to 8\,$M_{\odot}$ in increments of $0.5\,M_{\odot}$, and initial orbital periods covering 
$-0.3\leq \log_{10} (P_{\rm orb}^{\rm i}/\rm d)\leq 3.5$ in steps of 0.1. Following \citet{Nie2025}, the metallicity is set to be $Z=0.0142$.
Simulations are evolved until terminated by numerical limits, including: (1) logQ\_limit = 5, (2) gamma\_center\_limit = 1000, and (3) max\_number\_retries = 15000. Notably, we do not impose a maximum model number, as this could prematurely halt evolution before WD formation.

In our calculations, we adopt the mixing length theory of \citet{Bohm1958}, the Ledoux criterion for convection \citep{Ledoux1947}, and the default overshooting parameter in MESA. Nuclear reaction rates are taken from \citet{Cyburt2010} and \citet{Angulo1999}. Stellar winds are modeled using the Dutch prescription \citep{Glebbeek2009}, which incorporates wind models from  \citet{deJager1988}, \citet{Vink2001}, and \citet{Nugis2000} for different stellar types. Wind accretion onto the NS is treated via the Bondi-Hoyle mechanism \citep{Bondi1944}. We use the updated MT prescription of \citet{Marchant2021}, which based on the Roche-lobe radius fitting formula of \citet{Eggleton1983}. We assume that mass accretion onto the NS is limited by the Eddington rate, with the excess material being expelled from the binary and carrying away the specific orbital angular momentum of the NS. We note that some recent studies \citep[e.g.,][]{Li2021,Misra2025} suggest restricting the MT rate to sub-Eddington values, particularly for binaries forming NS$-$HeWD systems, to avoid excessive NS mass growth. In our simulations, the adopted Eddington-limited accretion yields a maximum NS mass of $\sim 2.6\,M_{\rm \odot}$, which is consistent with observed massive NSs but may overestimate accretion in certain MT phases.

Magnetic braking is implemented using the default prescription from \citet{Rappaport1983} in MESA, applied only to the donor star. And, magnetic braking is activated for binaries with donor masses $M_{\rm d}^{\rm i}\leq 1.5M_{\rm \odot}$ \citep[see e.g.,][]{Tauris1999,Podsiadlowski2002,Shao2020}. We note that recent work \citep{Van2019MN,Van2019} has shown that the default magnetic braking prescription may not accurately reproduce the observed properties of LMXBs, and a modified scheme has been proposed to better match observations. Furthermore, \citet{Y2025} explored the impact of this updated prescription on the formation of recycled pulsars. In our study, magnetic braking primarily affects systems undergoing stable MT, while our focus is on post-CE binaries where magnetic braking is inactive. Therefore, the choice of magnetic braking prescription is not expected to significantly alter our main conclusions regarding CE survivors.

We assume that a CE phase is triggered if the MT rate exceeds a threshold of $\dot{M}_{\rm CE} =0.1\,M_{\odot}\rm yr^{-1}$, distinct from the $\dot{M}_{\rm CE} =1\,M_{\odot}\rm yr^{-1}$ used in \citet{Nie2025}.
We adopt the standard energy formalism of \citet{Webbink1984} to model orbital evolution during CE phase, with the internal energy parameter set to $\alpha_{\rm th}=1.0$ \citep{Han1995}. We explore four values for the CE ejection efficiency $\alpha_{\rm CE}$ (3.0, 1.0, 0.3, and 0.1), which determines the fraction of orbital energy used to expel the donor's envelope.
During this phase, the envelope is removed at a constant mass-loss rate of $\dot{M}_{\rm CE}$. We define a high MT rate of $\dot{M}_{\rm high} = \dot{M}_{\rm CE}$ and a low MT rate of $\dot{M}_{\rm low} = 10^{-6}M_{\odot}\,\rm yr^{-1}$. If the donor radius $R_{\rm d}$ satisfies $(1-\delta)R_{\rm RL} < R_{\rm d} < R_{\rm RL}$, where $R_{\rm RL}$ is the Roche-lobe radius of the donor and $\delta$ is taken to be 0.02, the MT rate is interpolated between $\dot{M}_{\rm high}$ and $\dot{M}_{\rm low}$ following \citet{Marchant2021}. Most of the criteria for terminating the CE phase are consistent with those in \citet{Nie2025}. As demonstrated by \citet{Nie2025}, a binary may evolve into a CEDP during which a residual fraction of the hydrogen envelope is stably transferred to the NS via RLOF, making it a crucial phase for the formation of recycled pulsars.

\section{Result} \label{sec:result}

\subsection{The \texorpdfstring{$(M^{\rm i}_{\rm d}, P^{\rm i}_{\rm orb})$} P Parameter Space} \label{sec:3.1}
\begin{figure*}[htbp]
    \includegraphics[width=0.95\textwidth]{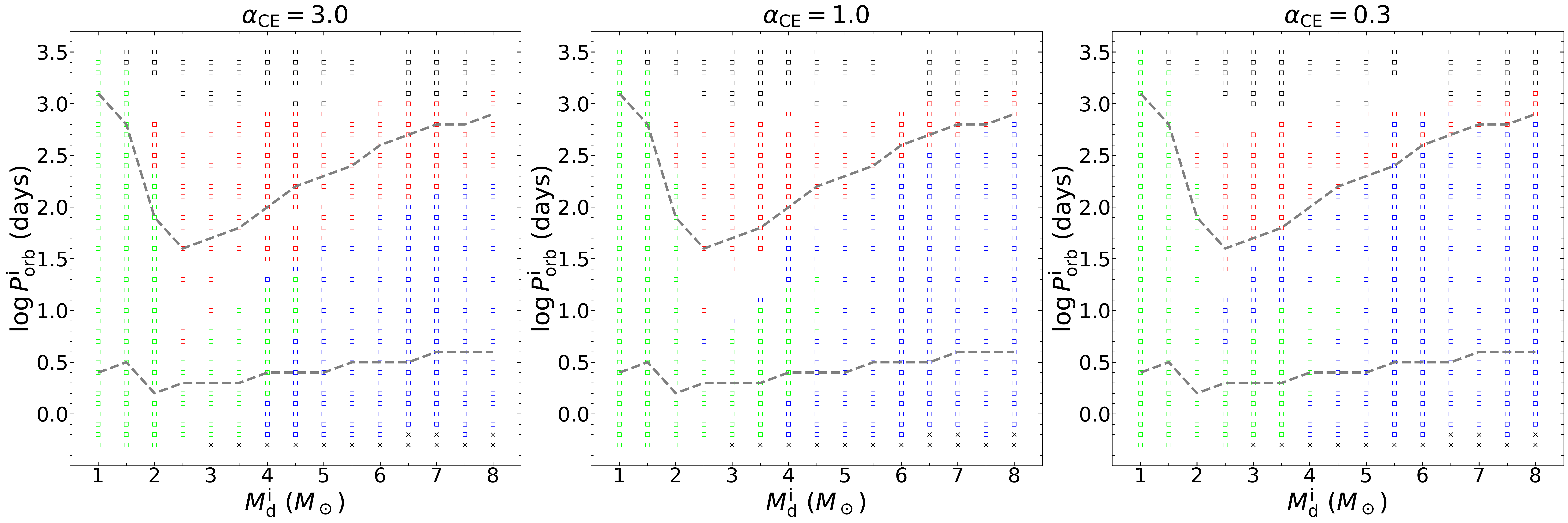}
    \centering
    \caption{The parameter space of initial donor mass versus orbital period ($M^{\rm i}_{\rm d}- P^{\rm i}_{\rm orb}$), illustrating the evolutionary fates of L/IMXBs with a $1.4\,M_\odot$ NS. The three panels correspond to CE ejection efficiencies of $\alpha_{\rm CE}=3.0$, 1.0 and 0.3. Green, blue, red, and black squares represent SMT binaries, CE mergers, CE survivors, and noninteracting binaries, respectively. Black crosses denote systems that are already Roche-lobe filling at the donor's ZAMS. The gray dashed curves separate the regimes of Case A, Case B, and Case C MT. The parameter space for CE survivors is shown to contract significantly with decreasing $\alpha_{\rm CE}$. Simulations for $\alpha_{\rm CE}=0.1$ (not shown) were attempted but most encountered numerical instabilities during the CE phase.}
    \label{1}
\end{figure*}

Figure \ref{1} presents the evolutionary outcomes of simulated binaries in the initial donor mass versus orbital period ($M^{\rm i}_{\rm d}-P^{\rm i}_{\rm orb}$) plane for different CE ejection efficiencies ($\alpha_{\rm CE}=3.0$, 1.0, and 0.3). The systems are categorized as follows:

\begin{enumerate}
\item[(1)] SMT binaries (green squares) - Systems undergoing stable MT (SMT) without initiating a CE phase.
\item[(2)] CE mergers (blue squares) - Systems that merge during CE evolution.
\item[(3)] CE survivors (red squares) - Systems that survive CE evolution and become close binaries with envelope-stripped donors.
\item[(4)] Noninteracting binaries (black squares) - Systems that do not experience any RLOF interaction.
\item[(5)] ZAMS RLOF binaries (black crosses) - Systems that are already Roche-lobe filling at the donor's ZAMS.
\end{enumerate}

SMT binaries occupy the parameter space  where $-0.3\leq \log_{10} (P_{\rm orb}^{\rm i}/\rm d)\leq 3.5$ and $M_{\rm d}^{\rm i}\leq 4.5\,M_{\rm \odot}$. For $M_{\rm d}^{\rm i}\leq 2\,M_{\rm \odot}$, SMT binaries cover a very wide range of initial orbital periods with $\log_{10} (P_{\rm orb}^{\rm i}/\rm d)\leq 3.5$, whereas for $M_{\rm d}^{\rm i}\geq 2.5\,M_{\rm \odot}$, they are confined to $\log_{10} (P_{\rm orb}^{\rm i}/\rm d)\leq 1.3$. Similar to \citet{Tauris2000a} and \citet{Shao2012}, the maximum $\log_{10} (P_{\rm orb}^{\rm i}/\rm d)$ of SMT binaries from our calculations decreases from 3.5 to 0.6 as $M_{\rm d}^{\rm i}$ from $1\,M_{\rm \odot}$ to $2.5\,M_{\rm \odot}$. 

CE survivors are found within $0.7\leq \log_{10} (P_{\rm orb}^{\rm i}/\rm d)\leq 3.1$ across all our adopted $\alpha_{\rm CE}$ values. The parameter space occupied by CE survivors shrinks notably as $\alpha_{\rm CE}$ decreases. For $\alpha_{\rm CE}=0.3$, when $M_{\rm d}^{\rm i}\geq 3\,M_{\rm \odot}$, the lower boundary of $\log_{10} P_{\rm orb}^{\rm i}$ and the upper gray dashed curve (distinguishing Case B and Case C MT) exhibit similar shapes. Above this boundary, all CE survivors are post-Case C binaries. The envelope binding energy in post-Case B donors is significantly larger than in post-Case C donors \citep[see also][]{Nie2025}. For $\alpha_{\rm CE}=3.0$ and $\alpha_{\rm CE}=1.0$, a substantial fraction of CE survivors at smaller $\log_{10} P_{\rm orb}^{\rm i}$ are post-Case B binaries.

\begin{figure*}[htbp]
    \includegraphics[width=0.95\textwidth]{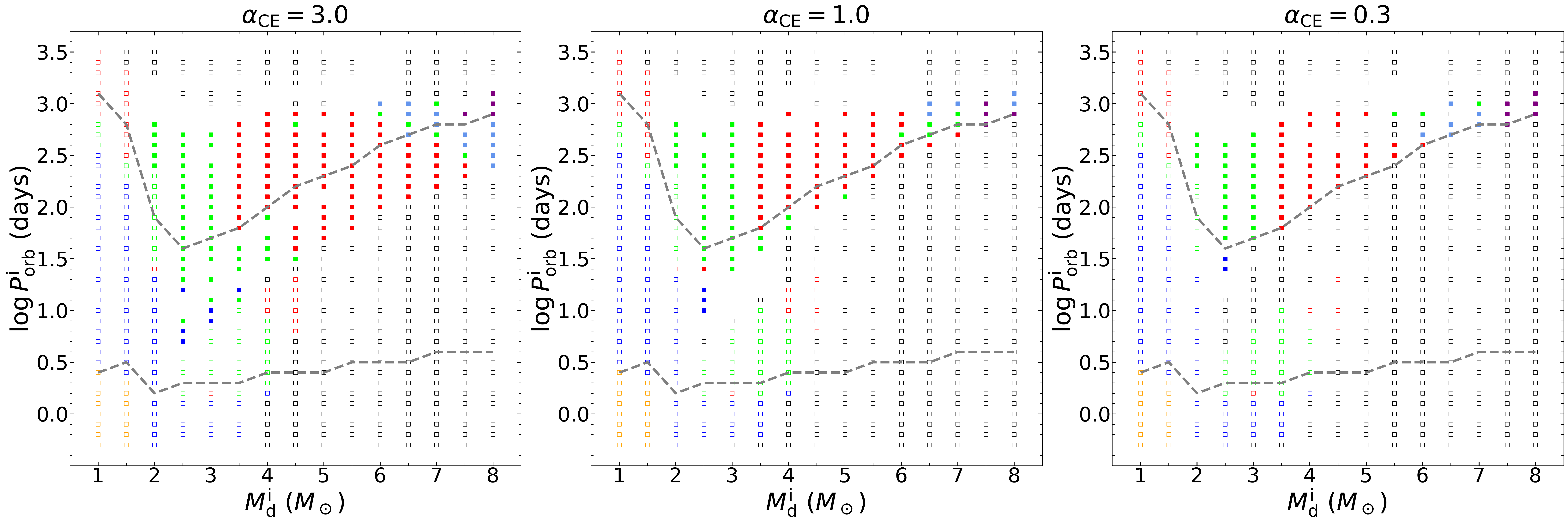}
    \centering
    \caption{Final fates of donor stars across the initial parameter space for $\alpha_{\rm CE}=3.0$, 1.0 and 0.3. The compact remnants are categorized as HeWDs (blue), HyWDs (green), COWDs (red), ONeWDs (azure), NSs (purple), or degenerate hydrogen-rich stars (orange). Black squares denote ZAMS RLOF binaries, CE mergers, and noninteracting binaries. Filled and open squares distinguish CE survivors and SMT binaries, respectively. The gray dashed curves separate the regimes of Case A, Case B, and Case C MT.} 
    \label{12}
\end{figure*}

\subsection{The \texorpdfstring{$(M^{\rm i}_{\rm d},\, P^{\rm i}_{\rm orb})$} P Parameter Space for Final Donor Fates}
\begin{figure*}[htbp]
    \centering
    \begin{minipage}{0.32\textwidth}
    \includegraphics[width=\linewidth]{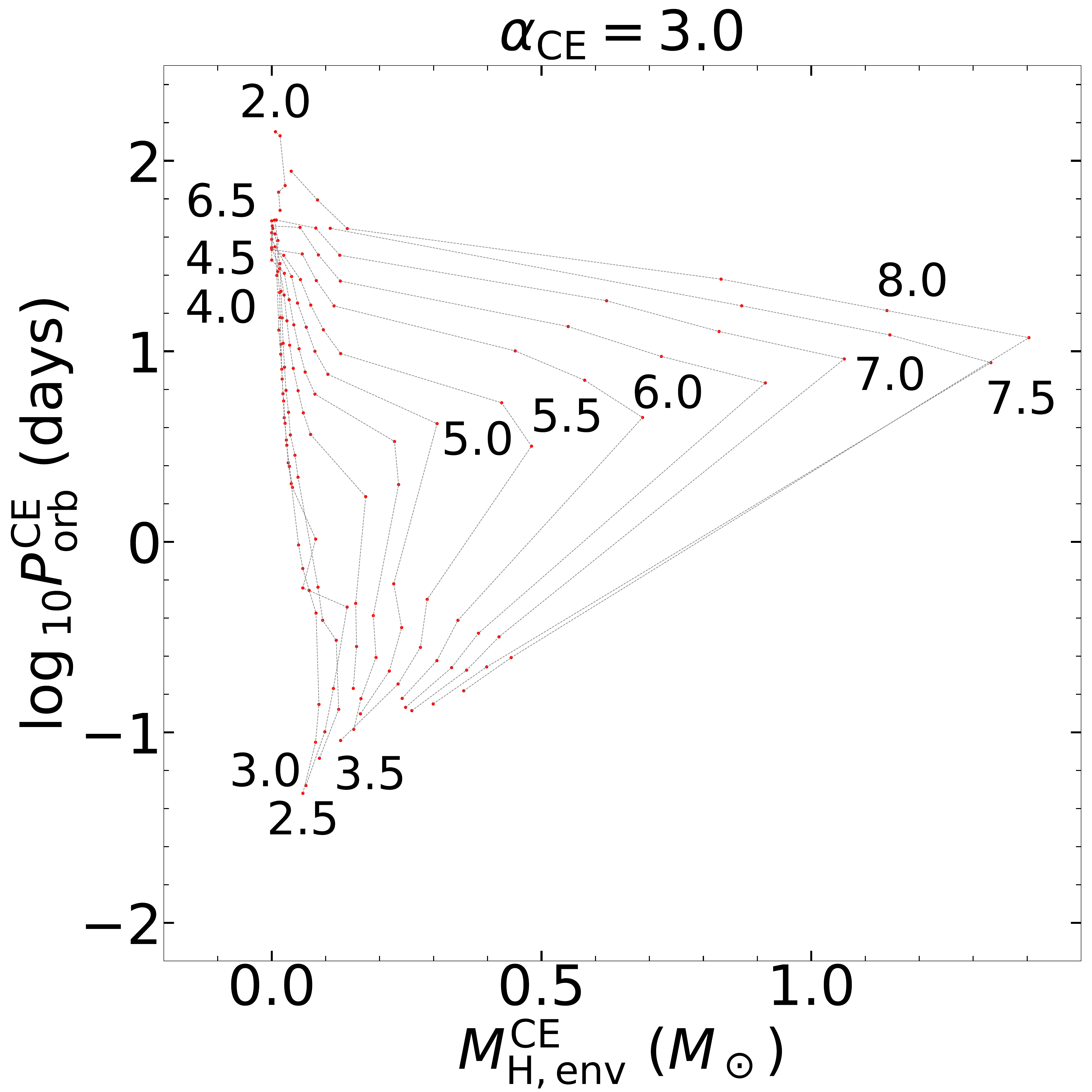}
    \end{minipage}
    \hfill
    \begin{minipage}{0.32\textwidth}
    \includegraphics[width=\linewidth]{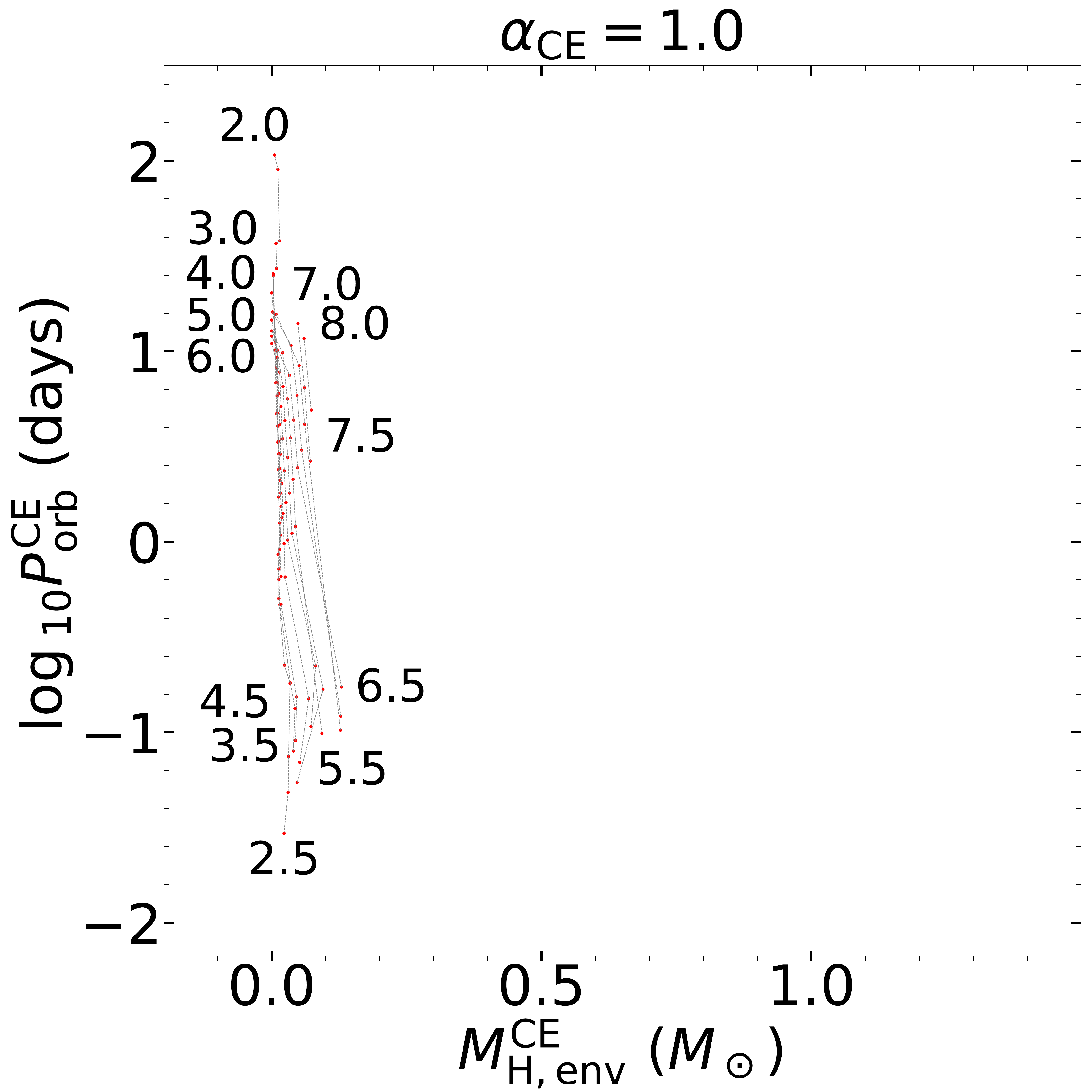}
    \end{minipage}
    \hfill
    \begin{minipage}{0.32\textwidth}
    \includegraphics[width=\linewidth]{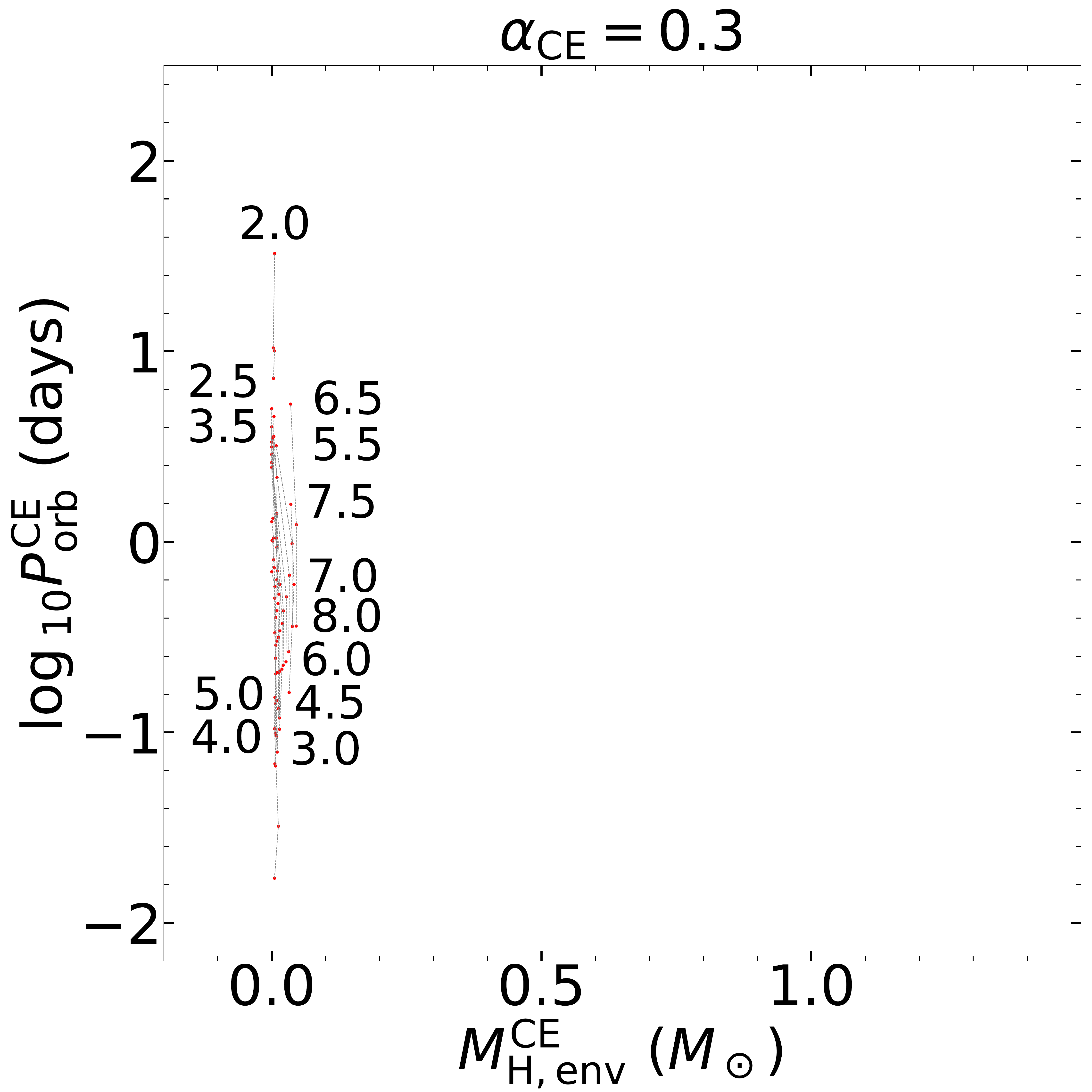}
    \end{minipage}
    \hfill
    \caption{Orbital period versus donor's hydrogen envelope mass ($P^{\rm CE}_{\rm orb}-M^{\rm CE}_{\rm H,env}$) at the termination of CE evolution. The three panels display the distributions for  $\alpha_{\rm CE}=3.0$, 1.0 and 0.3. The red dots represent all simulated systems that survived CE evolution. The number annotated next to each data cluster indicates the initial mass of the donor star. 
    }
    \label{4}
\end{figure*}

\begin{figure*}[htbp]
    \centering
    \begin{minipage}{0.32\textwidth}
    \includegraphics[width=\linewidth]{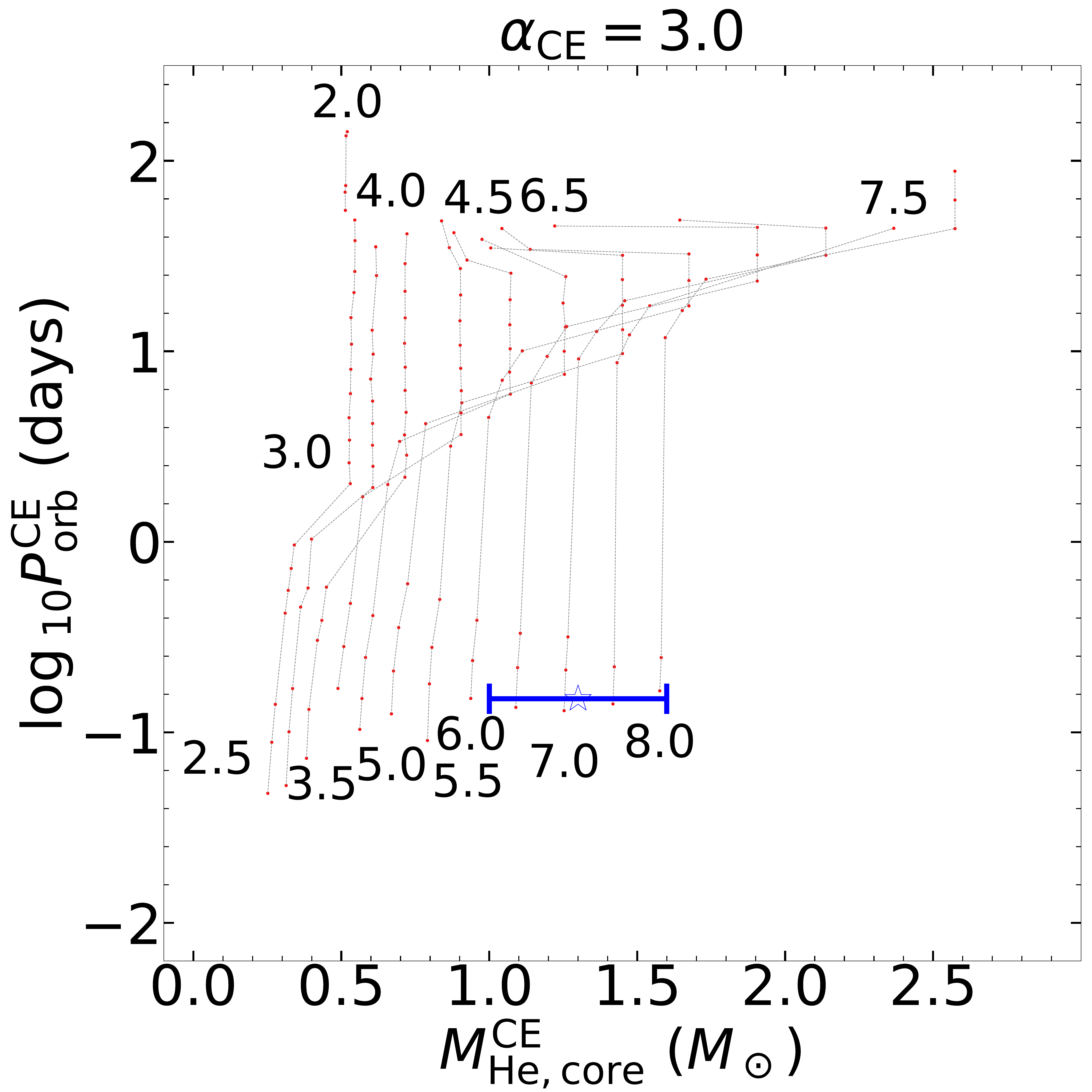}
    \end{minipage}
    \hfill
    \begin{minipage}{0.32\textwidth}
    \includegraphics[width=\linewidth]{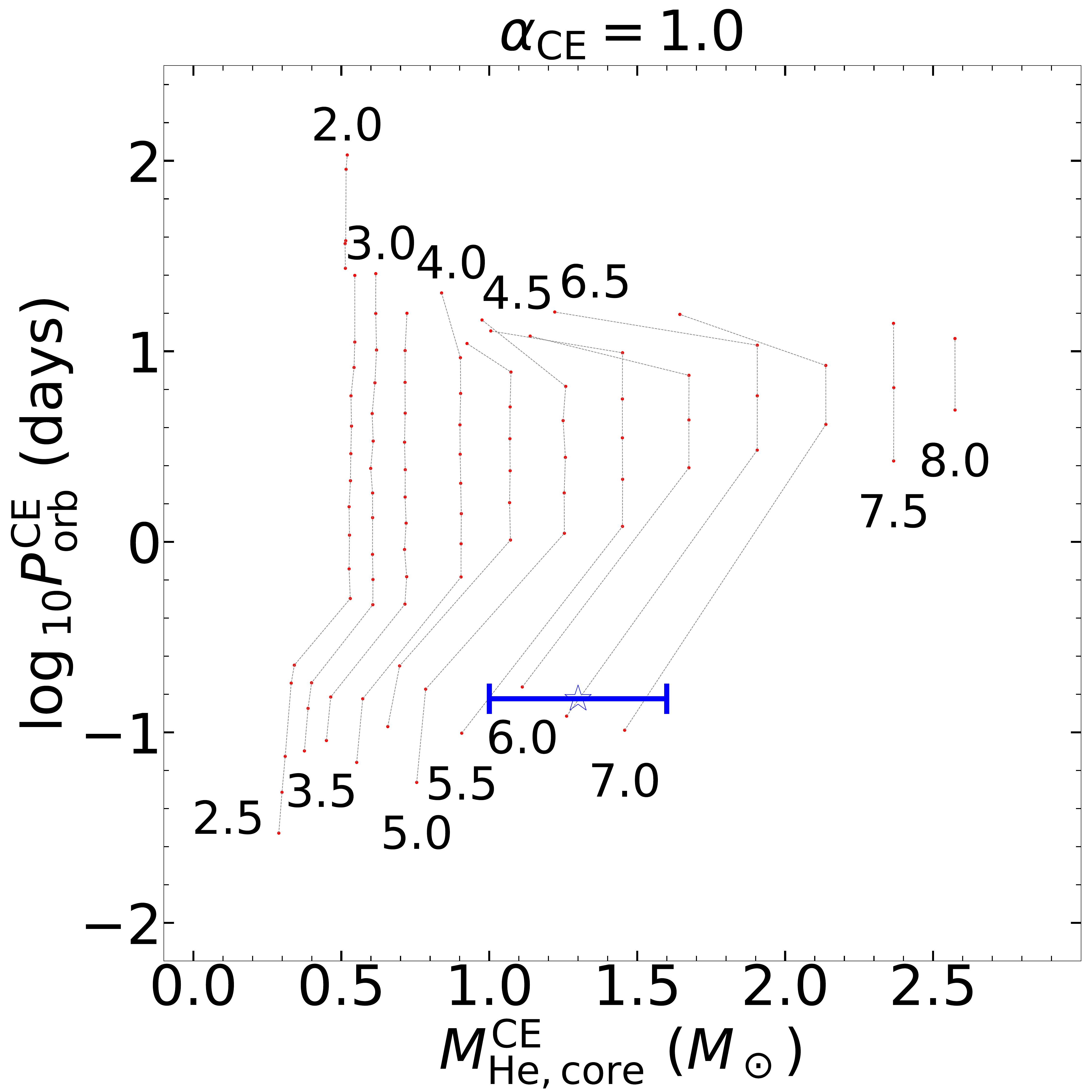}
    \end{minipage}
    \hfill
    \begin{minipage}{0.32\textwidth}
    \includegraphics[width=\linewidth]{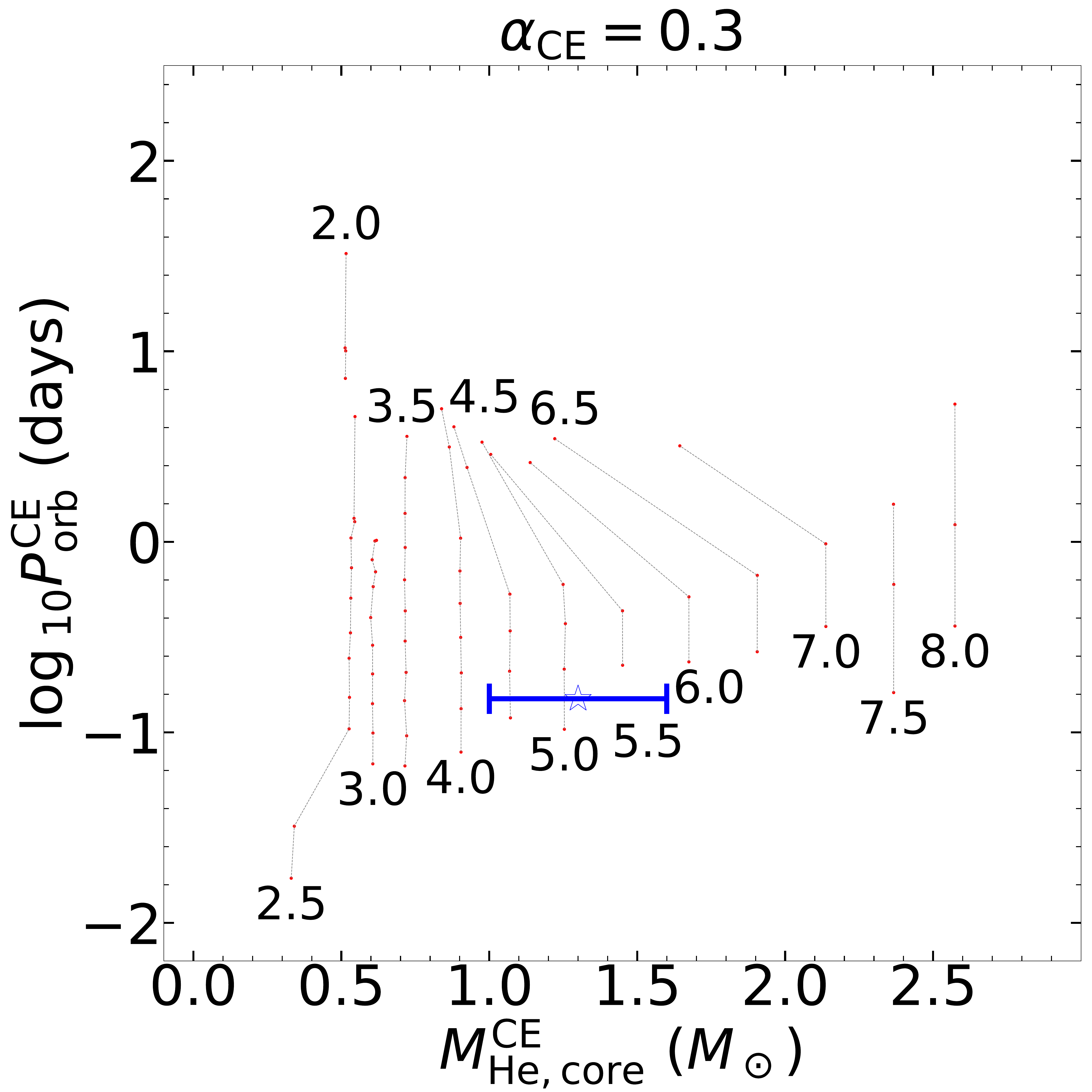}
    \end{minipage}
    \hfill
    \caption{Orbital period versus donor's He core mass ($\log_{10} P^{\rm CE}_{\rm orb}$,  $M^{\rm CE}_{\rm He,core}$) at the termination of CE evolution for $\alpha_{\rm CE}=3.0$, 1.0 and 0.3. Red dots denote all systems that survived CE evolution. The number annotated next to each data cluster indicates the initial donor mass. The blue star and horizontal error bar mark the central value and mass range, respectively, of the He star companion in PSR J1928+1815.}
    \label{6}
\end{figure*}

Figure \ref{12} shows the final evolutionary fates of the donor stars, classified as:

$\bullet$ NS$-$HeWD (blue squares): Donor evolves into a HeWD (CO core mass $<0.01\%$ of donor mass or no CO core).

$\bullet$ NS$-$HyWD (green squares): Donor becomes a hybrid WD (HyWD, CO core mass between $0.01\% $ and $ 95\%$ of donor mass).

$\bullet$ NS$-$COWD (red squares): Donor becomes a COWD (CO core mass $>95\%$ of donor mass).

$\bullet$ NS$-$ONeWD (azure squares): Donor becomes an ONeWD (mass fraction of elements heavier than oxygen in the core $>0.1$).

$\bullet$ NS$-$NS (purple squares): Donor evolves into an NS if its CO mass reaches $1.37M_{\rm \odot}$, forming a double NS system \citep[see also][]{Nie2025}.

$\bullet$ Converging systems (orange squares): Systems evolve with decreasing orbital periods until the hydrogen-rich donor becomes degenerate.

Overall, NS$-$COWD binaries are primarily the end fates of CE survivors, while NS$-$HeWD and NS$-$HyWD binaries are mainly produced by SMT binaries. The minimum initial donor mass ($M_{\rm d}^{\rm i}$) for forming NS$-$COWD binaries from CE survivors remains unchanged as $\alpha_{\rm CE}$ decreases from 3.0 to 0.3. Converging binaries appear where $-0.3\leq \log_{10} (P_{\rm orb}^{\rm i}/\rm d)\leq 0.4$ and $1\,M_{\rm \odot}\leq M_{\rm d}^{\rm i}\leq 1.5\,M_{\rm \odot}$. The donor fates for SMT binaries are generally consistent with \citet{Shao2012} and \citet{Misra2020}.

For SMT systems:

$\bullet$ NS$-$HeWD binaries form in the parameter-space region where $-0.3\leq \log_{10} (P_{\rm orb}^{\rm i}/\rm d)\leq 2.5$. 

$\bullet$ NS$-$HyWD binaries occupy $0.2 \leq \log_{10} (P_{\rm orb}^{\rm i}/\rm d)\leq 1$ for $2.5\,M_{\rm \odot}\leq M_{\rm d}^{\rm i}\leq 4M_{\rm \odot}$, and $1.5 \leq \log_{10} (P_{\rm orb}^{\rm i}/\rm d)\leq 2.8$ for $1\,M_{\rm \odot}\leq M_{\rm d}^{\rm i}\leq 2\,M_{\rm \odot}$. 

$\bullet$ NS$-$COWD binaries are found where $2.5\leq \log_{10} (P_{\rm orb}^{\rm i}/\rm d)\leq 3.5$ for $M_{\rm d}^{\rm i}\sim 1-1.5\,M_{\rm \odot}$, and  $0.8\leq \log_{10} (P_{\rm orb}^{\rm i}/\rm d)\leq 1.3$ for $M_{\rm d}^{\rm i}\sim 4\,M_{\rm \odot}-4.5\,M_{\rm \odot}$.

For CE survivors:

$\bullet$ NS$-$NS binaries are post-Case C binaries with $M_{\rm d}^{\rm i}=7.5\,M_{\rm \odot}$ or $8\,M_{\rm \odot}$. 

$\bullet$ NS$-$HyWD binaries primarily have $2\,M_{\rm \odot}\leq M_{\rm d}^{\rm i}\leq 3\,M_{\rm \odot}$, while NS$-$COWD binaries span $3.5\,M_{\rm \odot}\leq M_{\rm d}^{\rm i}\leq 7.5\,M_{\rm \odot}$. 

$\bullet$ NS$-$ONeWD binaries exist for $6\,M_{\rm \odot}\leq M_{\rm d}^{\rm i}\leq 8\,M_{\rm \odot}$ at relatively longer $P_{\rm orb}^{\rm i}$. For $6\,M_{\rm \odot}\leq M_{\rm d}^{\rm i}\leq 7\,M_{\rm \odot}$, they are exclusively post-Case C binaries across all $\alpha_{\rm CE}$. For $\alpha_{\rm CE}=3.0$, NS$-$ONeWD binaries with $7.5\,M_{\rm \odot}\leq M_{\rm d}^{\rm i}\leq 8\,M_{\rm \odot}$ are post-Case B binaries. A unique case occurs for $\alpha_{\rm CE}=1.0$, where post-Case C binaries with $M_{\rm d}^{\rm i}=8\,M_{\rm \odot}$ and $3\leq \log_{10} (P_{\rm orb}^{\rm i}/\rm d)\leq 3.1$ form an ONeWD instead of an NS due to the mass loss via RLOF of the CO core during CEDP and subsequent stellar wind.

$\bullet$ NS$-$HeWD binaries from CE survivors appear in the region where $2.5\,M_{\rm \odot}\leq M_{\rm d}^{\rm i}\leq 3.5\,M_{\rm \odot}$ and $0.7\leq \log_{10} (P_{\rm orb}^{\rm i}/\rm d)\leq 1.5$ for all $\alpha_{\rm CE}$.

\subsection{Classifications of Binaries Experiencing Various Post-CE Evolutionary Phases} \label{sec:3.3}

In Figure \ref{4}, we present the distributions of systems at CE termination in the orbital period ($ P_{\rm orb}^{\rm CE}$) vs. donor's hydrogen-envelope mass ($M_{\rm H,env}^{\rm CE}$) plane. Wider-orbit CE survivors generally originate from wider initial orbits. The choice of $\alpha_{\rm CE}$ significantly affects the post-CE orbital period distribution, with $P_{\rm orb}^{\rm CE}$ ranging from $\sim 0.04-158$ days ($\alpha_{\rm CE}=3.0$) to $\sim 0.016-32$ days ($\alpha_{\rm CE}=0.3$).

In the case of $\alpha_{\rm CE}=3.0$, similar to the classifications of \citet{Nie2025}, post-Case C binaries have $P_{\rm orb}^{\rm CE} \sim 2-158$ days and $M_{\rm H,env}^{\rm CE}\lesssim 0.2\,M_\odot$. Post-Case B binaries, experienced a long duration \citep[$>1000\,$yr,][]{Nie2025} of CE evolution, have $P_{\rm orb}^{\rm CE} \lesssim 1\,$day and $M_{\rm H,env}^{\rm CE}\lesssim 0.5\,M_\odot$. In contrast, post-Case B binaries, experienced a short duration (typically a few years) of CE evolution, have $P_{\rm orb}^{\rm CE} \gtrsim 1\,$day and $M_{\rm H,env}^{\rm CE}\lesssim 1.4\,M_\odot$.
In the case of $\alpha_{\rm CE}=1.0$, post-Case C binaries occupy $M_{\rm H,env}^{\rm CE}\lesssim 0.1\,M_{\rm \odot}$ and post-Case B binaries with $P_{\rm orb}^{\rm CE}\lesssim 0.25\,$days occupy $M_{\rm H,env}^{\rm CE}\lesssim 0.2M_{\rm \odot}$. In the case of $\alpha_{\rm CE}=0.3$, most CE survivors are post-Case C binaries with $M_{\rm H,env}^{\rm CE}\lesssim 0.1\,M_{\rm \odot}$.

\begin{figure*}[htbp]
    \includegraphics[width=0.8\textwidth]{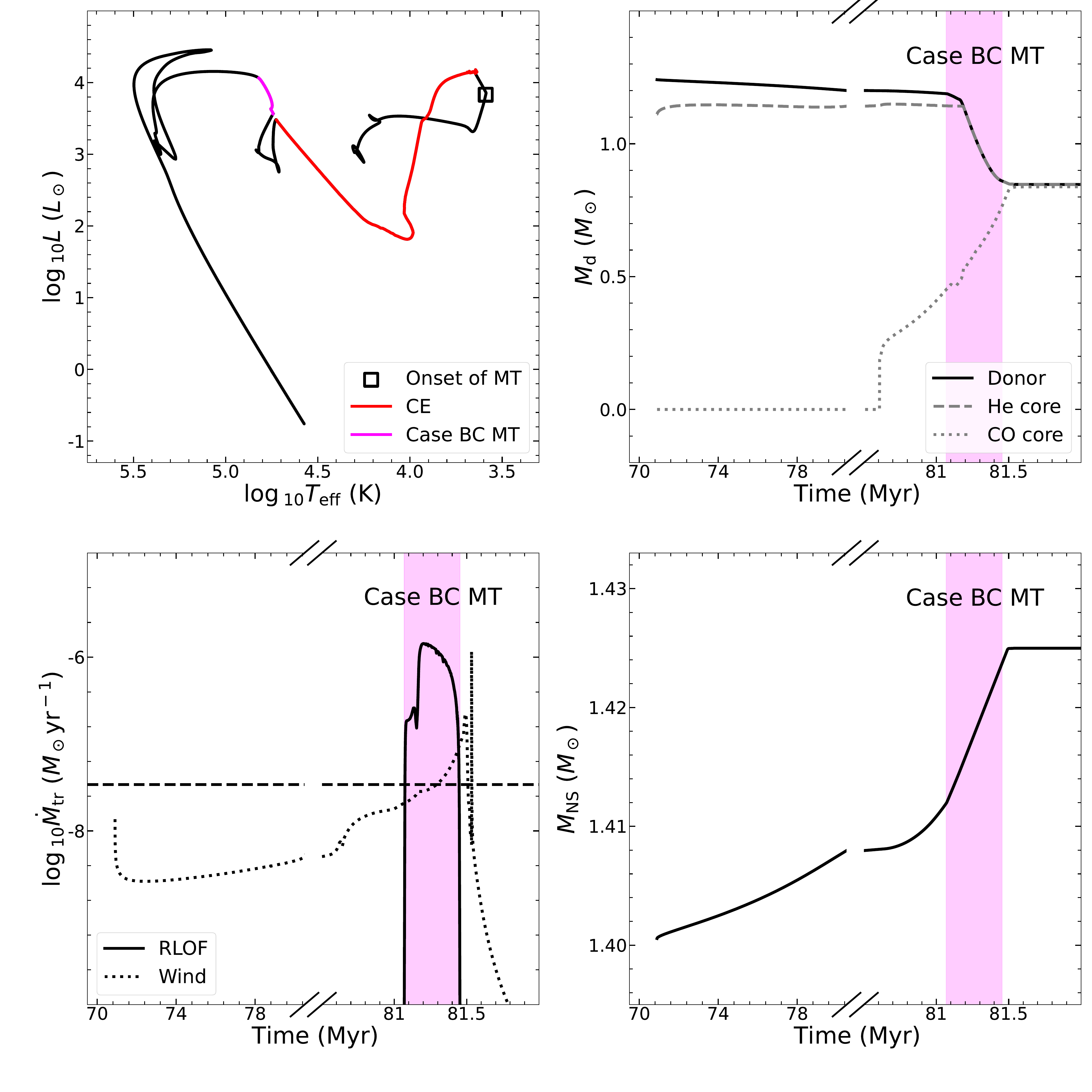}
    \centering
    \caption{Evolutionary tracks for a binary system with a $1.4M_{\odot}$ NS and a $6\,M_{\odot}$ donor in an initial 316-day orbit, computed with $\alpha_{\rm CE}=1.0$. Upper left: Hertzsprung-Russell diagram for the donor star. The black square marks the onset of MT. The red and magenta curves correspond to the binary undergoing a CE phase and the subsequent Case BC MT phase, respectively. Upper right: Evolution of the donor star mass (solid curve), He core mass (dashed curve), and CO core mass (dotted curve) after the CE phase. Lower left: MT rate via RLOF as a function of time. The black dashed line indicates the Eddington accretion limit. Lower right: Mass accretion history of the NS, showing the total mass gained over time.}
    \label{Observation1}
\end{figure*}

\begin{figure*}[htbp]
    \includegraphics[width=0.8\textwidth]{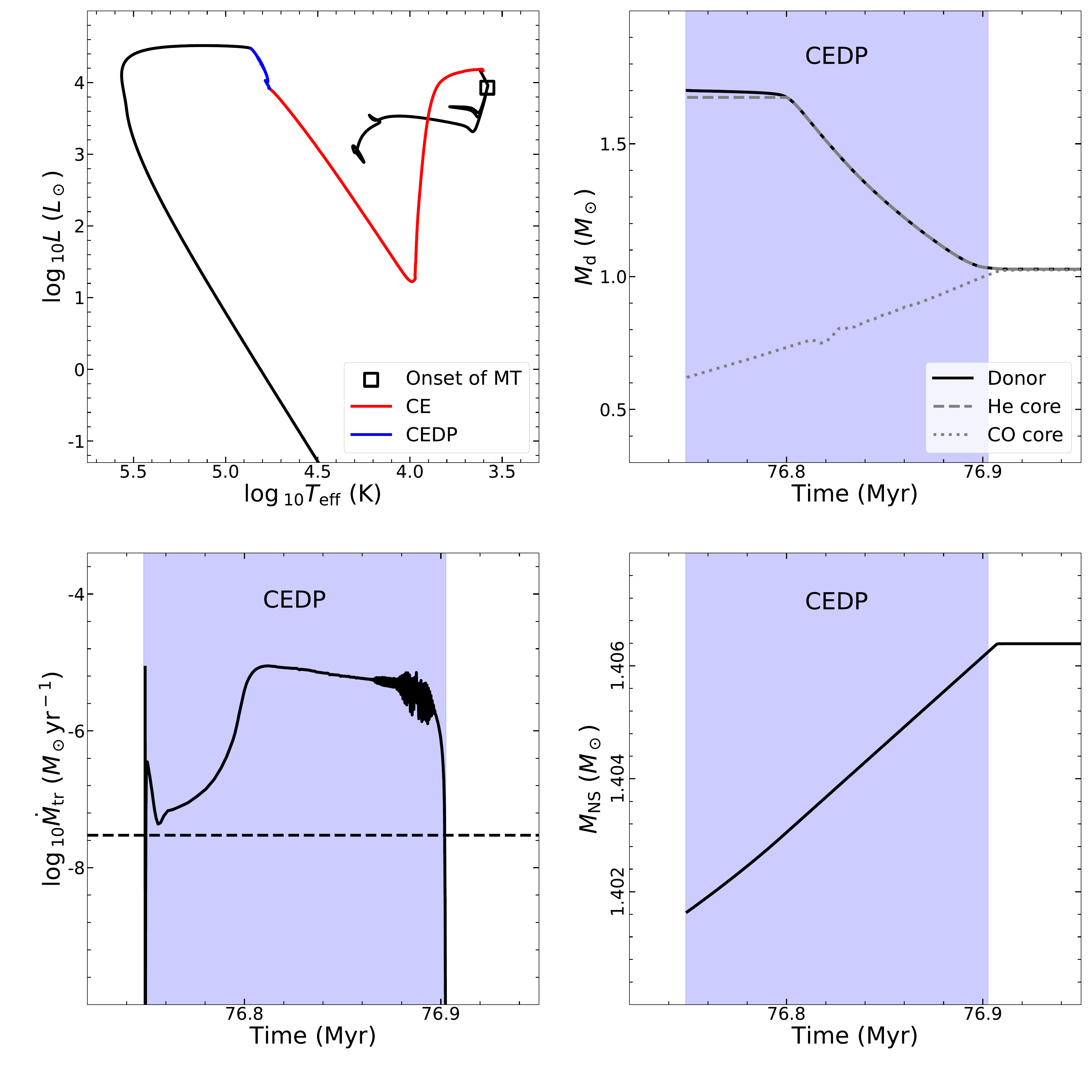}
    \centering
    \caption{An alternative evolutionary track for forming a system like PSR J1928+1815, from a binary with a $1.4M_{\odot}$ NS and a $6M_{\odot}$ donor in a wider initial orbit of 398 days, computed with $\alpha_{\rm CE}=0.3$. Upper Left: Hertzsprung-Russell diagram of the donor star. The black square marks the onset of MT. The red and blue curves correspond to the binary undergoing a CE phase and the subsequent CEDP, respectively. Upper Right: Evolution of the donor star mass (solid curve), He core mass (dashed curve), and CO core mass (dotted curve) after the CE phase. Lower Left: MT rate via RLOF as a function of time. The black dashed line indicates the Eddington accretion limit. Lower Right: Mass accretion history of the NS, showing the total mass gained over time.}
    \label{Observation2}
\end{figure*}

\begin{figure*}[htbp]
    \centering
    \begin{minipage}{0.32\textwidth}
    \includegraphics[width=\linewidth]{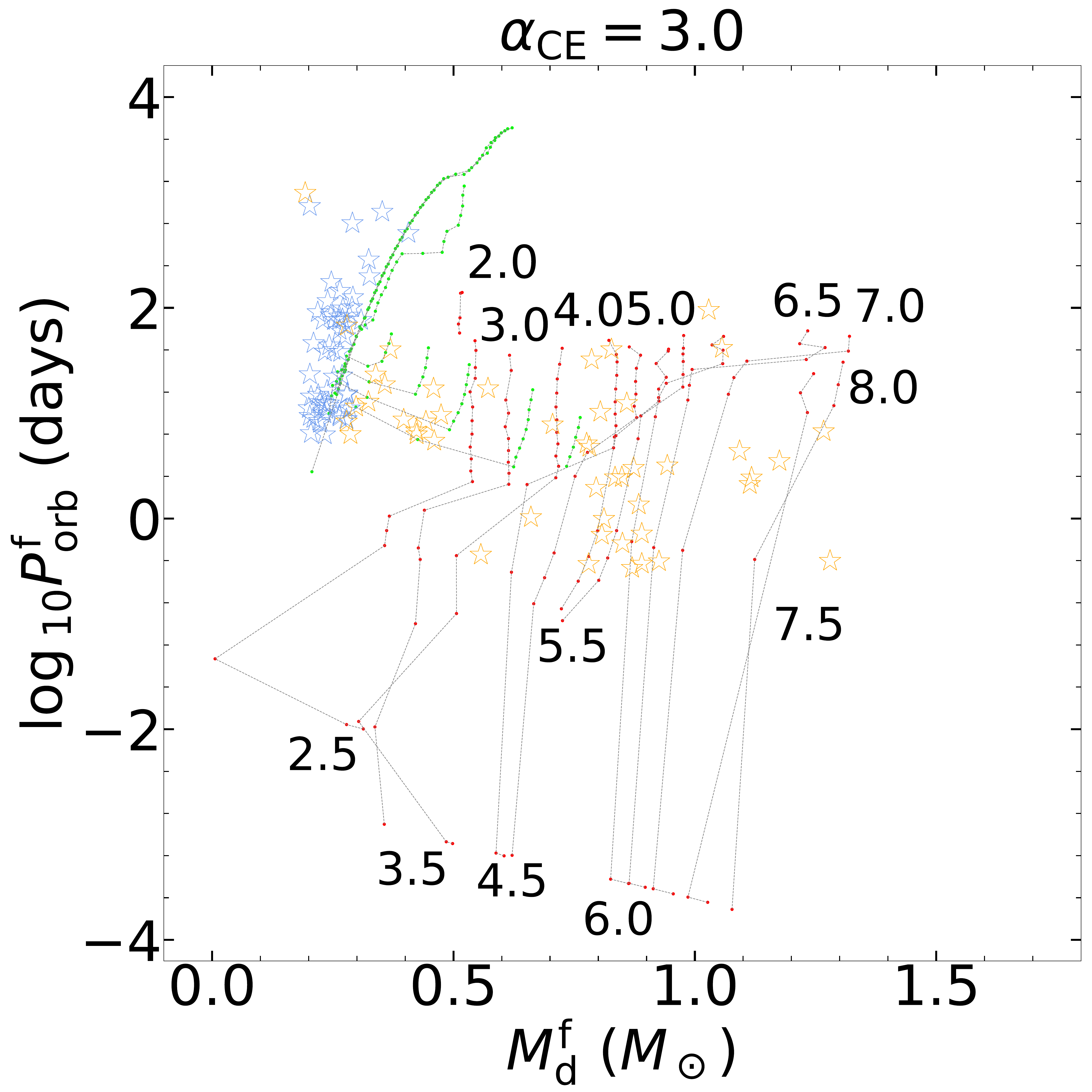}
    \end{minipage}
    \hfill
    \begin{minipage}{0.32\textwidth}
    \includegraphics[width=\linewidth]{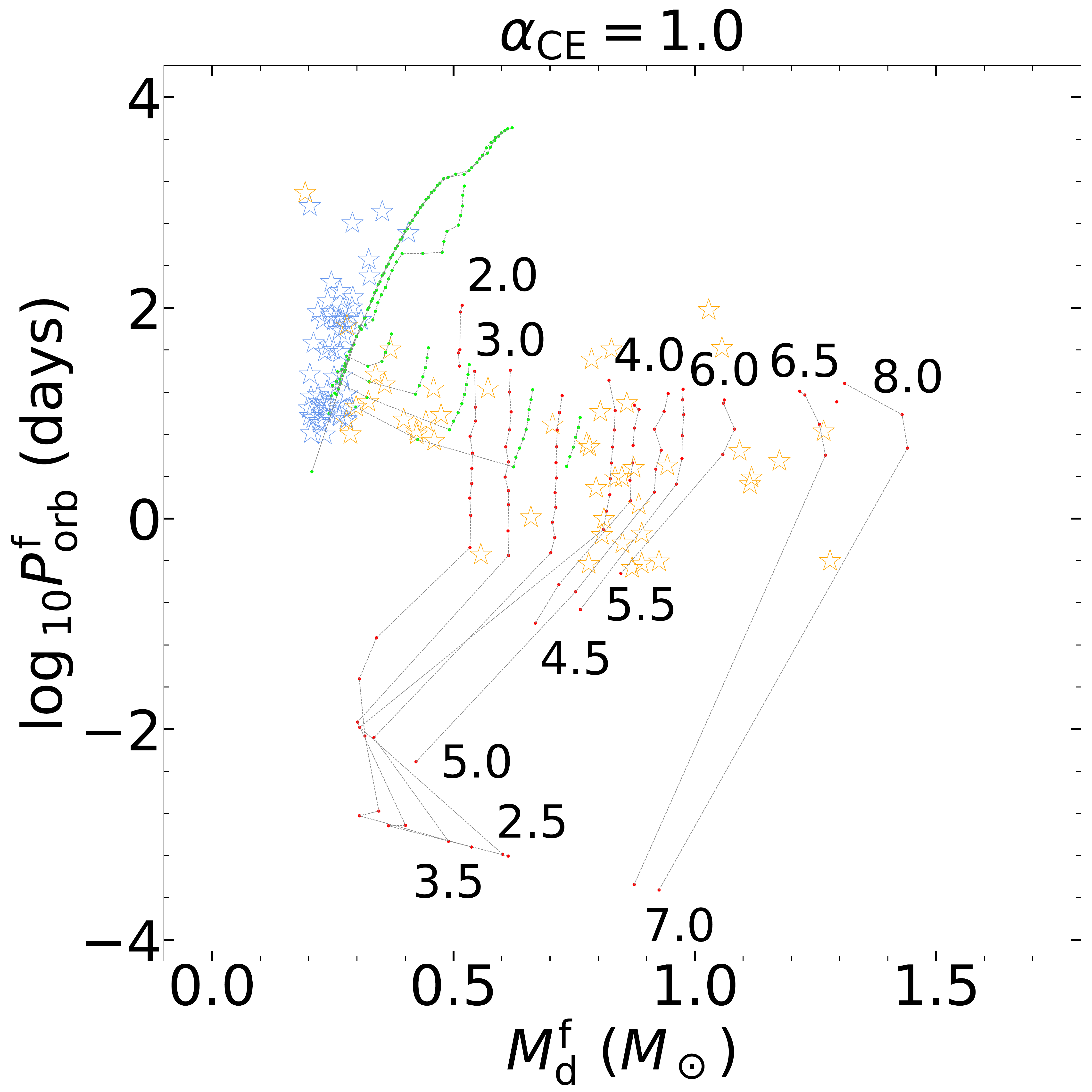}
    \end{minipage}
    \hfill
    \begin{minipage}{0.32\textwidth}
    \includegraphics[width=\linewidth]{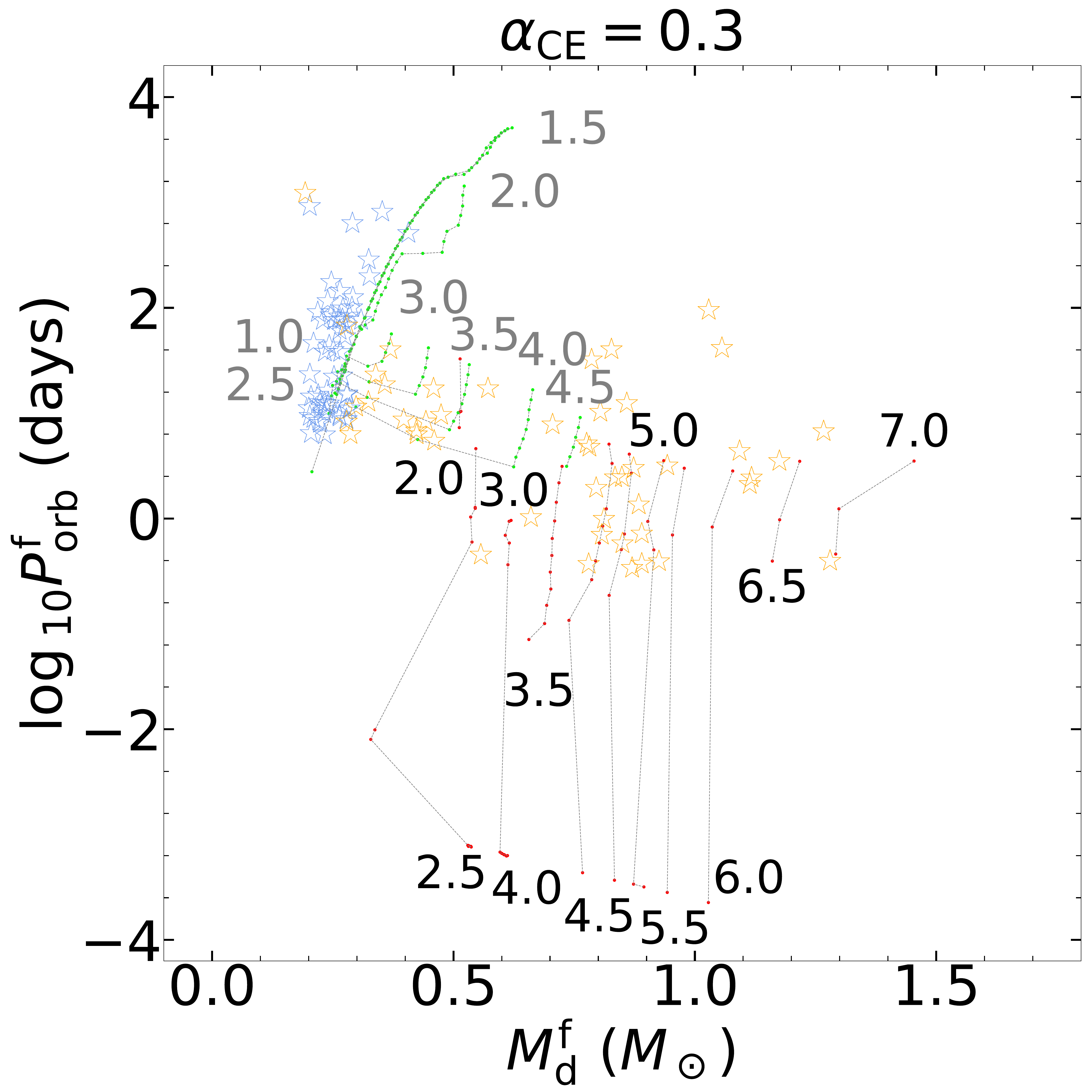}
    \end{minipage}
    \hfill
    \caption{Final orbital period verses final donor mass ($\log_{10} P^{\rm f}_{\rm orb}$, $M^{\rm f}_{\rm d}$) for all simulated NS$-$WD binaries from both CE survivor (red dots) and SMT (green dots) channels, for CE ejection efficiencies $\alpha_{\rm CE}=3.0$, 1.0 and 0.3. The number next to each curve indicates the initial donor mass. Observed NS$–$HeWD binaries and NS$–$COWD/ONeWD binaries are over-plotted as blue and orange stars, respectively, with data primarily from the ATNF pulsar catalogue \citep{Manchester2005} and \citet{Yang2025WD}. For each observed binary, the $M_{\rm d}^{\rm f}$ is the minimum WD mass assuming an orbital inclination of $90^{\rm \circ}$.}
    \label{7}
\end{figure*}

\subsection{Binary Evolution of the PSR J1928+1815 System}

\subsubsection{Parameters at CE Termination}
Figure \ref{6} presents orbital period ($P_{\rm orb}^{\rm CE}$) versus donor's He core mass ($M_{\rm He,core}^{\rm CE}$, including the CO core and its He envelope) at CE termination. CE survivors consistently have $M_{\rm He,core}^{\rm CE}\lesssim 2.6\,M_\odot$.

Post-Case C binaries with a specific $M_{\rm d}^{\rm i}$ have longer orbital periods and nearly identical $M_{\rm He,core}^{\rm CE}$, which is $\sim 0.2-1\,M_{\odot}$ more massive than those of post-Case B binaries. An exception occurs for post-Case C binaries with $4\,M_{\rm \odot}\leq M_{\rm d}^{\rm i}\leq 7\,M_{\rm \odot}$ and relatively long $P_{\rm orb}^{\rm i}$, where $M_{\rm He,core}^{\rm CE}$ is less massive due to convective envelope expansion dredging He into the envelope before CE onset. All binaries with $M_{\rm d}^{\rm i}=2\,M_{\rm \odot}$ are post-Case C binaries with $M_{\rm He,core}^{\rm CE}\sim 0.5\,M_{\rm \odot}$.

\subsubsection{Properties of PSR J1928+1815}

The eclipsing millisecond pulsar PSR J1928+1815, likely a post-CE system, orbits a $\sim 1.0-1.6\,M_{\rm \odot}$ He star with a period of $\sim0.15$ days \citep{Yang2025}. The He star does not exceed its Roche lobe. Based on the pulsar's spin period, it is estimated to have accreted at least $0.01\,M_{\rm \odot}$ during the CE phase.

\subsubsection{Evolutionary Tracks for PSR J1928+1815}
For $\alpha_{\rm CE}=3.0$ and 1.0, the parameters of post-Case B binaries are able to match PSR J1928+1815 well. Figure \ref{Observation1} shows an example track for a system with a $1.4\,M_{\rm \odot}$ NS and a $6\,M_{\rm \odot}$ donor in a 316-day orbit ($\alpha_{\rm CE}=1.0$). After CE evolution, stellar wind mass loss reduces the donor by $\sim 0.05\,M_{\rm \odot}$, with the NS accreting $\lesssim 0.01\,M_{\rm \odot}$ over $\sim 10^{\rm 7}\,$yr. If the NS accreted the required $0.01\,M_{\rm \odot}$  during the CE phase, the system becomes a detached binary with the donor slightly underfilling its Roche lobe - conditions favorable for observing an eclipsing pulsar for $\sim 10^{\rm 7}\,$yr. The donor mass remains at $\sim 1.2M_{\rm \odot}$, consistent with the observed companion. We propose that PSR J1928+1815 may be in this stage. Subsequent Case BC MT ($\sim 81.07\,$Myr) increase the orbit to $\sim 0.3\,$days and reduces the donor mass to $\sim 0.85M_{\rm \odot}$, inconsistent with observations.

For $\alpha_{\rm CE}=0.3$, post-Case C binaries can match the observations. Figure \ref{Observation2} shows a track for the same mass but a wider orbit of $P_{\rm orb}^{\rm i}=398\,$days ($\alpha_{\rm CE}=0.3$). After CE evolution and the CEDP, the donor loses $\sim 0.7\,M_{\rm \odot}$ of its He envelope, and the NS accretes $\lesssim 0.007\,M_{\rm \odot}$, leaving a $\sim 1.02M_{\rm \odot}$ donor consistent with observations. After CEDP, gravitational wave radiation  shrinks the orbit to $\sim 0.15\,$days in $\sim 3.65\,$Gyr, matching the observed period. An eclipsing millisecond pulsar would be observable if the NS gained sufficient mass during the CE phase.

In our simulations, systems where the NS accretes a sufficient amount of material ($\gtrsim 0.01\,M_{\rm \odot}$) are typically characterized by donor masses below $1\,M_{\rm \odot}$. This leads us to favor the scenario in which PSR J1928+1815 accreted at least $0.01\,M_{\rm \odot}$ during the CE phase, consistent with the constraints derived by \citet{Yang2025}. Recently, \citet{Guo2025} suggested that Case BA/BB MT during post-CE evolution could also form millisecond pulsars similar to PSR J1928+1815. However, their simulations did not include the complete evolutionary pathways beginning from IMXBs through a CE phase. Moreover, \citet{Deng2025} proposed that NSs could accrete enough mass to become millisecond pulsars resembling PSR J1928+1815 if super-Eddington accretion occurs prior to CE evolution. Additionally,  \citet{Gong2025} obtained deep infrared photometry of PSR J1928+1815 and found no evidence for a He star companion, suggesting instead that the companion may be an inflated WD in a black widow type system. Their results are compatible with our post–Case C binary model for $\alpha_{\rm CE}=0.3$ (Figure \ref{Observation2}), in which the binary evolves into an NS$–$COWD system as the orbit shrinks to the observed period.

\subsection{Comparison with Observed binary pulsars}\label{Sec:3.6}

\begin{figure*}[htbp]
    \centering
    \begin{minipage}{0.32\textwidth}
    \includegraphics[width=\linewidth]{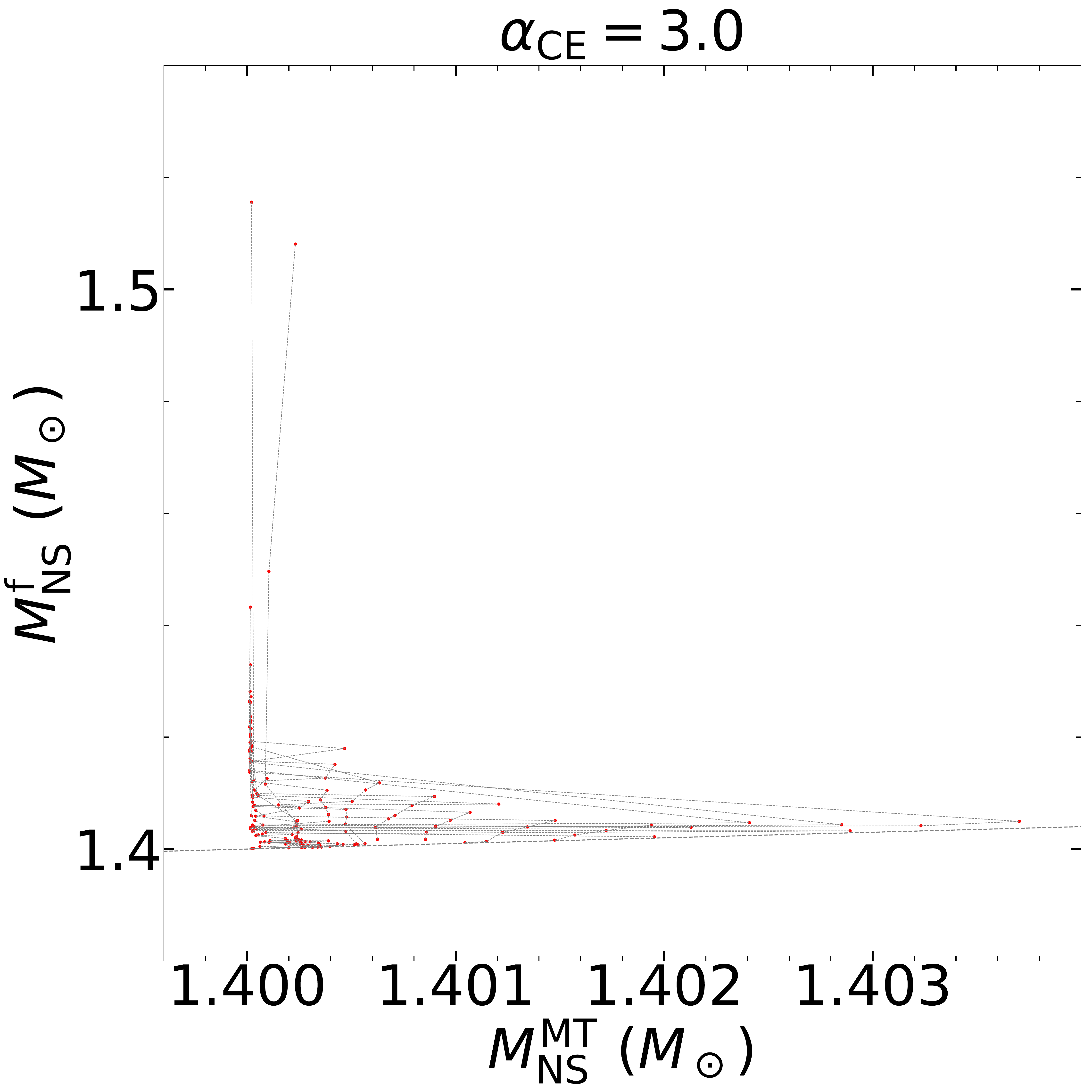}
    \end{minipage}
    \hfill
    \begin{minipage}{0.32\textwidth}
    \includegraphics[width=\linewidth]{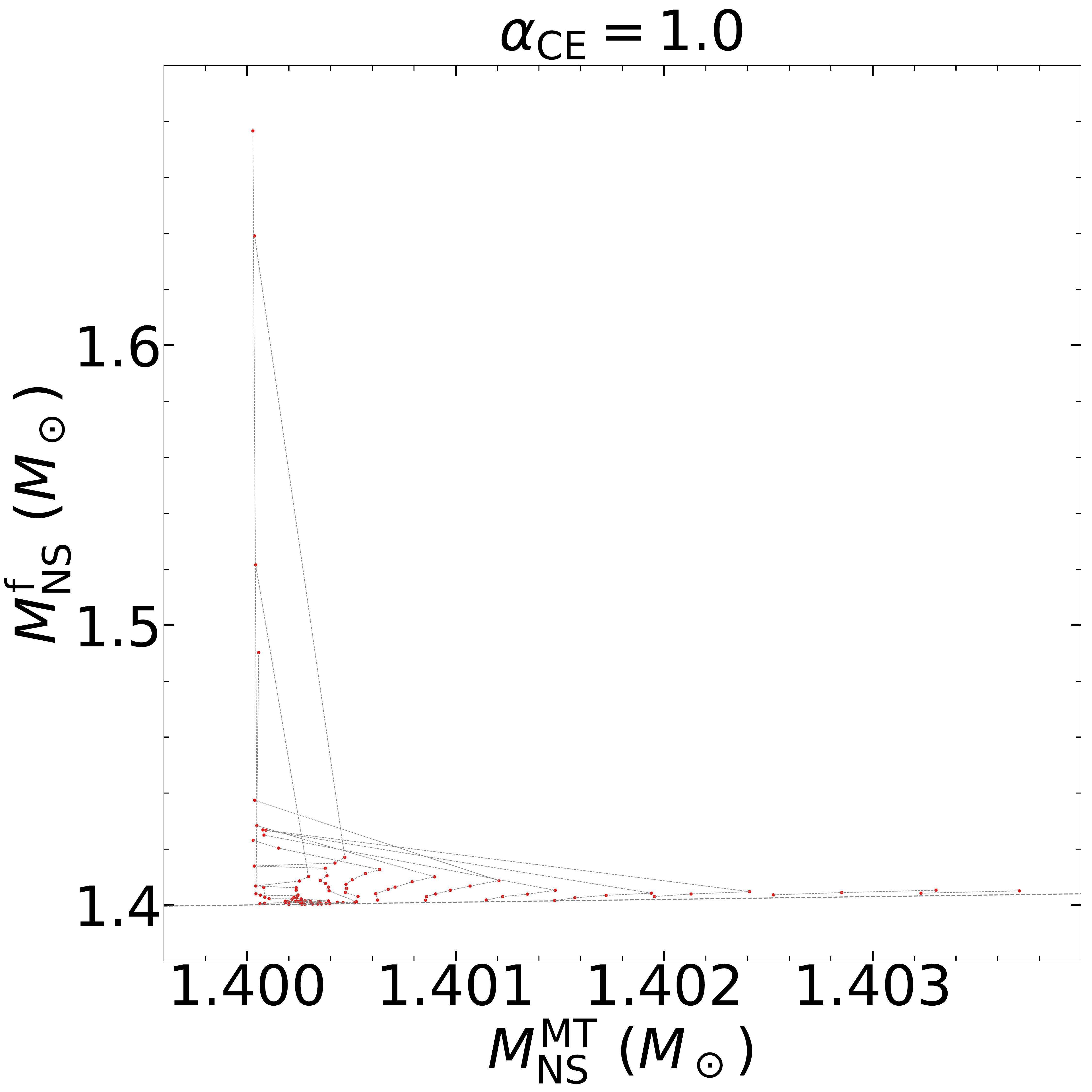}
    \end{minipage}
    \hfill
    \begin{minipage}{0.32\textwidth}
    \includegraphics[width=\linewidth]{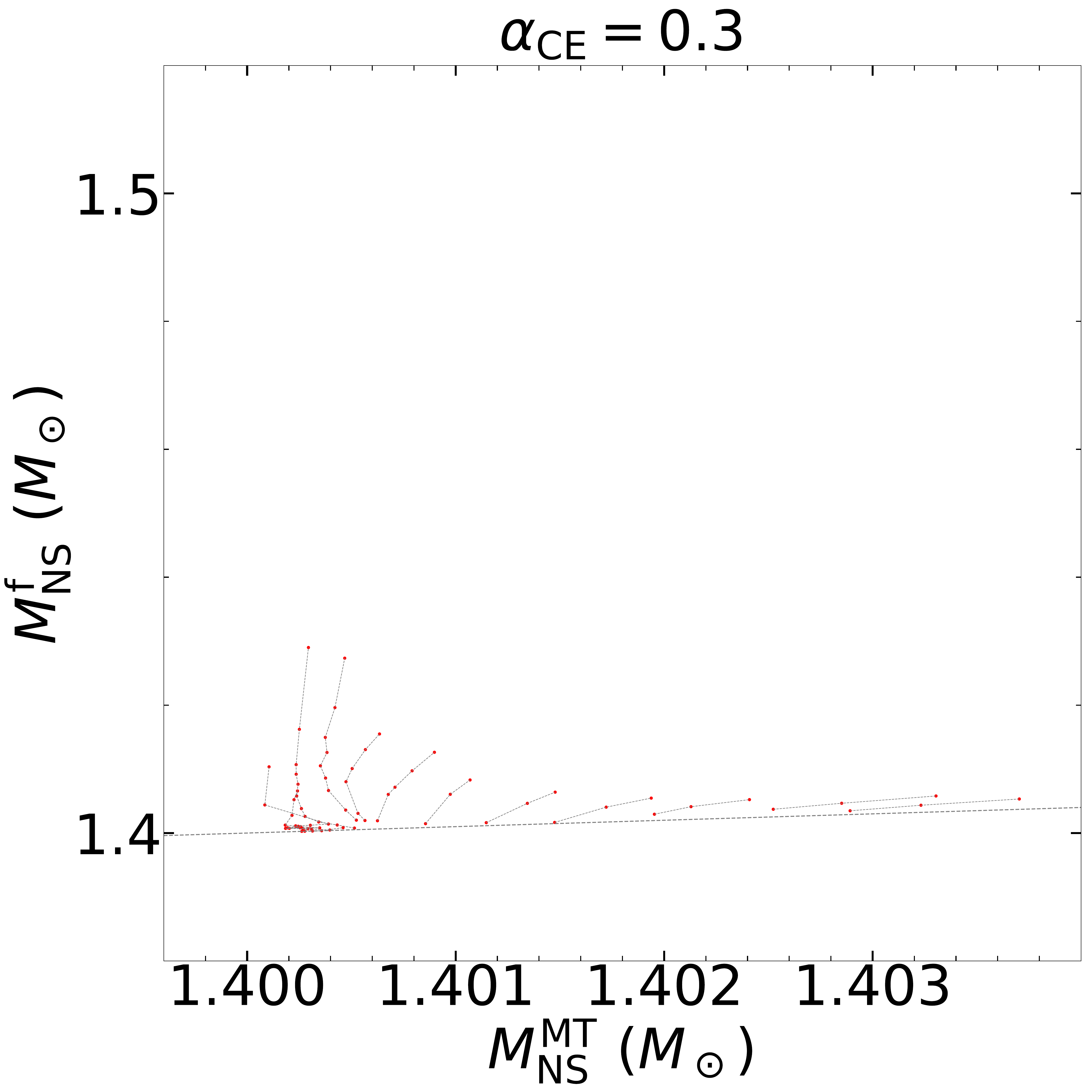}
    \end{minipage}
    \hfill
    \caption{Final NS mass ($M^{\rm f}_{\rm NS}$) as a function of its mass at the onset of the first RLOF ($M^{\rm MT}_{\rm NS}$), under the assumption of $\alpha_{\rm CE}=3.0$, 1.0 and 0.3. The red dots represent systems that survived CE evolution. The dashed curve in each panel indicates the relation of $M^{\rm f}_{\rm NS}=M^{\rm MT}_{\rm NS}$.}
    \label{9}
\end{figure*}

Figure \ref{7} shows final orbital period ($P_{\rm orb}^{\rm f}$) versus final donor mass ($M_{\rm d}^{\rm f}$) for simulated NS$-$WD binaries from both CE survivors and SMT channels, compared with observed systems. 

For SMT binaries with $M_{\rm d}^{\rm i}\sim 2-4.5\,M_{\rm \odot}$, donors from early Case B MT systems are more massive than those from Case A MT systems. For SMT binaries with $M_{\rm d}^{\rm i}\leq 1.5\,M_{\rm \odot}$, $P_{\rm orb}^{\rm f}$ increases with $M_{\rm d}^{\rm f}$ \citep[see also][]{Tauris2000a}. 

For CE survivors with a specific $M_{\rm d}^{\rm i}$, post-Case C donors have nearly identical $M_{\rm d}^{\rm f}$, $\sim 0.2-1\,M_{\rm \odot}$ more massive than post-Case B donors with shorter $P_{\rm orb}^{\rm f}$. As $\alpha_{\rm CE}$ decreases from 3.0 to 0.3, the maximum $P_{\rm orb}^{\rm f}$ for CE survivors decreases significantly.  The minimum $P_{\rm orb}^{\rm f}$ is $\sim 10^{\rm -4}\,$days across all $\alpha_{\rm CE}$, insensitive to $\alpha_{\rm CE}$ due to gravitational wave radiation works. NS$-$COWD binaries have $0.5\,M_{\rm \odot}\lesssim M_{\rm d}^{\rm f}\lesssim 1.1\,M_{\rm \odot}$.

Observed NS$-$WD binaries \citep{Manchester2005,Yang2025WD} are over-plotted. One system (orange) with reported companion mass  $\sim 0.2M_{\rm \odot}$ and  $P_{\rm orb}^{\rm f}\sim 1200\,$days is an outlier, its actual mass is likely $\sim 0.6\,M_{\rm \odot}$ \citep{Koester2001}. Simulated systems with $M_{\rm d}^{\rm i}\leq 2\,M_{\rm \odot}$ roughly match observations. However, observed binaries with $P_{\rm orb}^{\rm f}\gtrsim 60\,$days systematically deviate, having lighter WD companions than our models \citep[see also][]{Shao2012}.

SMT binaries experienced Case B MT can match some observed NS$-$WD binaries with less massive COWDs. Although these are classified as NS$-$HyWD binaries, their donors have substantial CO cores ($\sim 0.3-0.6\,M_{\rm \odot}$). 


As CE survivors, post-Case C binaries provide a better match to observed NS–WD systems harboring more massive COWDs. The agreement with observations is strongly influenced by the choice of CE ejection efficiency.
It should be noted that some simulated tracks terminate prematurely due to numerical limitations, which complicates a direct comparison between the final evolutionary states and the observed sample. Under the assumption that observed binaries with massive COWD/ONeWD companions indeed underwent CE evolution, they should correspond to post-CE systems. 
For $\alpha_{\rm CE}=0.3$, observed systems with orbital periods of $\sim 10-100\,$days and companion masses $\sim 0.8-1.3\,M_{\rm \odot}$ are not matched by our models, since systems with initial donor masses of $4\,M_{\rm \odot}\leq M_{\rm d}^{\rm i}\leq 6.5\,M_{\rm \odot}$ only reach a maximum post-CE orbital period of $\sim 5\,$days. Even for $\alpha_{\rm CE}=1.0$, the maximum orbital period only extends to $\sim 20\,$days for the same initial mass range, still falling short of the $100\,$days seen in observations. Therefore, based on these observational constraints, a higher CE efficiency of $\alpha_{\rm CE}=3.0$ is clearly preferred in simulating the CE phase of IMXBs. 

We also note that NSs in CE survivors may accrete a modest amount of material \citep{MacLeod2015}, potentially in conjunction with jet activity \citep{Papish2015}. Such processes are not included in the present simulations but could contribute to envelope ejection and influence the final binary properties, suggesting an important direction for future work.
\subsection{A Study on the Recycling process}

In Figure \ref{9}, we present the final NS mass ($M^{\rm f}_{\rm NS}$) versus the NS mass at the first RLOF onset ($M^{\rm MT}_{\rm NS}$) for all CE survivors. Note that the initial NS mass is set to be $1.4\,M_\odot$. The maximum accreted mass is $\sim 0.28\,M_{\rm \odot}$ across all $\alpha_{\rm CE}$. In post-Case C binaries, NSs accrete $\lesssim 0.004\,M_{\rm \odot}$ via stellar wind \citep{deJager1988} before RLOF/CE and up to $\sim 0.01\,M_{\rm \odot}$ during CEDPs (lasting $\lesssim 0.5$~Myr).

In post-Case B binaries, stellar wind accretion prior to RLOF, modeled according to  \citet{deJager1988}, is found to be negligible. In the case of $\alpha_{\rm CE}=3.0$, for the systems with $M_{\rm d}^{\rm i}\leq 3\,M_{\odot}$, experienced a long CE duration, NSs accrete primarily via RLOF during Case BA/BB/BC MT phases, with one case even reaching $\sim 0.1\,M_{\odot}$ of accretion as the donor evolves into a WD. In contrast, those with short CE durations accrete mainly during CEDPs. For the systems with $M_{\rm d}^{\rm i}\geq 3.5\,M_{\odot}$, mass gain from RLOF (CEDPs and Case BB/BC MT phases) is comparable to wind capture.
Overall, NSs accrete $\sim 0.01-0.1\,M_{\rm \odot}$ in close post-Case B binaries and $\lesssim 0.01\,M_{\rm \odot}$ in wide systems, potentially becoming mildly recycled pulsars \citep{Tauris2012}. 
For $\alpha_{\rm CE}=0.3-1.0$, NSs in post-Case B binaries accrete $\sim 0.002-0.28\,M_{\rm \odot}$ via RLOF in Case BC MT phases as well as from the \citet{Nugis2000} stellar wind.

\section{Conclusion} \label{sec:conclusion}

We have employed extensive grids of MESA simulations, incorporating recently updated treatments for CE evolution, to investigate the evolutionary pathways of L/IMXBs leading to binary pulsars. Our models are initiated with a ZAMS donor star and a $1.4\,M_{\rm \odot}$ NS, exploring initial orbital periods spanning $-0.3\leq\log_{10} (P_{\rm orb}^{\rm i}/\rm d)\leq 3.5$ and initial donor masses ranging from $1\,M_{\odot}$ to $8\,M_{\odot}$. Exploring the effect of CE ejection efficiencies ($\alpha_{\rm CE} = 3.0$, 1.0 and 0.3), we summarize our principal findings as follows.

The parameter space occupied by CE survivors in the $M^{\rm i}_{\rm d}-\log_{10} P^{\rm i}_{\rm orb}$ diagram contracts significantly as $\alpha_{\rm CE}$ decreases. For systems undergoing SMT, our derived initial parameter space aligns well with previous studies such as \citet{Tauris2000a}, and the final evolutionary outcomes of our simulated binaries are largely consistent with results from \citet{Misra2020}.

Our calculations indicate that NS$-$COWD binaries primarily originate from CE survivors with initial donor masses $M_{\rm d}^{\rm i}\geq 3.5\, M_{\rm \odot}$, yielding final WD masses in the range $0.5\,M_{\rm \odot}\lesssim M_{\rm d}^{\rm f}\lesssim 1.1\,M_{\rm \odot}$. The minimum initial donor mass required to form a COWD from a CE survivor remains unchanged as $\alpha_{\rm CE}$ decreases from 3.0 to 0.3. For $\alpha_{\rm CE}=1.0$ and 0.3, NS$-$ONeWD binaries form exclusively via post-Case C binaries when $6\,M_{\rm \odot}\leq M_{\rm d}^{\rm i}\leq 8\,M_{\rm \odot}$, whereas for $\alpha_{\rm CE}=3.0$, they can result from both post-Case B and post-Case C channels within the same mass range. NS$-$NS systems form exclusively as post-Case C binaries from donors with 
$M_{\rm d}^{\rm i}\sim 7-8\,M_{\rm \odot}$.


All CE survivors in our simulations are characterized by a He core mass in the range $0.25\,M_{\odot}\lesssim M_{\rm He,core}^{\rm CE}\lesssim 2.6\,M_{\odot}$ at the termination of the CE phase.
For the millisecond pulsar PSR J1928+1815, which has an observed companion mass of $1.0-1.6\,M_{\rm \odot}$ and an orbital period of $\sim 0.15\,$days \citep{Yang2025}, we identify two plausible formation channels. The observed parameters are consistent with those of a post-Case B binary shortly after CE evolution but before subsequent Case BC MT (see Figure \ref{Observation1}), a configuration expected to manifest as an eclipsing millisecond pulsar for 
$10^{\rm 7}\,$yr. Alternatively, a post-Case C binary following the CEDP (see Figure \ref{Observation2}) can also match the observational constraints.

The minimum final orbital period ($P_{\rm orb}^{\rm f}$) in our simulations is approximately $10^{\rm -4}-10^{\rm -3}\,$days for all $\alpha_{\rm CE}$, dictated by gravitational wave radiation. Consequently, the minimum $P_{\rm orb}^{\rm f}$ shows little sensitivity to the choice of $\alpha_{\rm CE}$. Comparisons with the observed population of NS–WD binaries reveal that the parameters of SMT binaries align well with observed systems harboring HeWDs or less massive COWDs. Conversely, post-Case C binaries better match observed NS–WD systems with more massive CO WD companions. As $\alpha_{\rm CE}$ increases, the calculated parameters of CE survivors show better agreement with observed binaries having relatively long orbital periods. These results indicate that a higher CE ejection efficiency ($\alpha_{\rm CE}=3.0$) provides the best match to the observational data.

NSs in post-Case B binaries primarily gain mass ($\sim0.01-0.1\,M_\odot$) through RLOF during CEDPs and Case BA/BB/BC MT phases, as well as via stellar wind accretion \citep{Nugis2000}, potentially becoming mildly recycled pulsars. In contrast, NSs in post-Case C binaries accrete predominantly via stellar wind \citep{deJager1988} before the onset of RLOF/CE and during CEDPs, with total accretion typically $\lesssim 0.01\,M_\odot$. Our simulations demonstrate that both scenarios can produce recycled pulsars, consistent with observational evidence.


\begin{acknowledgments}
This work was supported by the National Key Research and Development Program of China (grant Nos 2021YFA0718500 and 2023YFA1607902), the Natural Science Foundation of China (Nos 12041301, 12121003, and 12373034), and the Strategic Priority Research Program of the Chinese Academy of Sciences (Grant No. XDB0550300). All input files to reproduce our results are available for download from Zenodo at \dataset[doi:10.5281/zenodo.18041567]{https://doi.org/10.5281/zenodo.18041567}.
\end{acknowledgments}




\bibliography{reference}{}
\bibliographystyle{aasjournal}



\end{document}